\documentclass[letter,aps,prlt,twocolumn,showpacs]{revtex4}
\usepackage{graphicx,color}
\usepackage{times}
\usepackage{booktabs}
\newcommand{\mrm}[1]{\mathrm{#1}}
\newcommand{\tsc}[1]{\textsc{#1}}
\newcommand{\ttt}[1]{\texttt{#1}}
\newcommand{\parm}[1]{\ttt{\footnotesize #1}}
\newcommand{\Nch}{\ensuremath{N_{\mrm{ch}}}}
\newcommand{\pT}{\ensuremath{p_\perp}}
\newcommand{\avgpT}{\ensuremath{\left<\pT\right>}}
\newcommand{\pTofNch}{\ensuremath{\avgpT(\Nch)}}
\newcommand{\TeV}{\,\mbox{Te\kern-0.2exV}}
\newcommand{\GeV}{\,\mbox{Ge\kern-0.2exV}}
\newcommand{\MeV}{\,\mbox{Me\kern-0.2exV}}
\newcommand{\keV}{\,\mbox{ke\kern-0.2exV}}
\newcommand{\eV}{\,\mbox{e\kern-0.2exV}}

\newcommand{\tbar}{\bar{t}}

\begin{document}

\title{Non-perturbative QCD Effects and the Top Mass at the Tevatron}
\author{Peter Skands$^1$\footnote{skands@fnal.gov}, Daniel Wicke$^2$\footnote{Daniel.Wicke@physik.uni-wuppertal.de}}
\affiliation{$^1$Theoretical Physics, MS106, Fermi National Accelerator
  Laboratory, Batavia, IL 
  60510-0500, USA\\
$^2$Bergische Universit\"at, Fachbereich C, Physik,
 Gaustr.~20, 42097 Wuppertal, Germany}

\begin{abstract}
We present a new, universally applicable toy model of colour
reconnections in hadronic final states. The model is based on
hadronising strings and has one free parameter.
We next present an implementation of this model in the \tsc{Pythia}
event generator and provide several parameter sets (`tunes'), 
constrained by fits to Tevatron minimum-bias data.
Finally, we consider the sensitivity of a simplified top mass
analysis to these effects, in exclusive semi-leptonic top events 
at the Tevatron. A first attempt at
isolating the genuine non-perturbative effects gives an estimate 
of order $\delta m_{\mathrm{top}}
\sim \pm0.5\GeV$ from non-perturbative uncertainties,
and a further $\delta m_\mathrm{top} \sim \pm1\GeV$ from shower effects. 
\end{abstract}
\pacs{
12.38.-t, % Quantum Chromodynamics
13.85.Hd, % Inelastic scattering: many-particle final states
13.87.Fh  % fragmentation into hadrons
~;~FERMILAB-PUB-06-340-T}
\maketitle

\section{Introduction}

With increasing statistics and improved analysis techniques,
a truly precise measurement of the top quark mass 
now seems feasible at the Tevatron experiments, reaching a final
uncertainty at or below 1.5\GeV~\cite{Brubaker:2006xn}. This is all
the more impressive given that the top mass is a highly non-trivial
observable, involving both jets and leptons.  
Moreover, it furnishes an important motivation  to 
reconsider which theoretical aspects are relevant, at the 1\GeV\ 
level, and whether they are sufficiently well under
control. Ultimately, this question will also be relevant for a range
of proposed high-precision measurements at the LHC. 

In particular for hadronic final states, a sophisticated
array of corrections are applied to the experimental raw data
before  the actual observable is
evaluated \cite{Bhatti:2005ai,Brubaker:2006xn,D'Hondt:2006nc}. 
Due to the increasingly advanced procedures  
mandated by high precision, it is not straightforward to predict how
uncertainties in the modelling affect the final answer; instead, 
dedicated studies are required to establish whether 
theoretical models are sufficiently well constrained
and/or whether modified measurement strategies could ultimately be
more fruitful. 

On the theoretical side, techniques for consistent matching between
perturbative parton showers and fixed-order calculations have 
been improved and generalised in recent years (for reviews see e.g.\ 
\cite{Mangano:2005dj,Hoche:2006ph,Sjostrand:2006su}), 
with some work focusing specifically on top 
production
\cite{Norrbin:2000uu,Frixione:2003ei,Hamilton:2006ms,Mangano:2006rw}.   
The structure of the underlying event (UE)
has also received increasing attention
\cite{Affolder:2001xt,Field:2005sa,Field:2005yw,Abdullin:2006me,Acosta:2006bp,Group:2006rt}, 
with theoretical developments here focusing on resummations of 
multiple perturbative interactions (MPI)
\cite{Sjostrand:1987su,Butterworth:1996zw,Sjostrand:2004pf,Sjostrand:2004ef}.
Non-perturbative aspects, on the other hand,  
still suffer from being hard to quantify, hard to test, and hard
to calculate. In this study, we focus on one particular such source of
uncertainty: colour reconnection effects in the final state. 

We begin by briefly discussing some general aspects of
colour reconnections, including the role 
they already play in current descriptions of hadron collisions. 
We next present several explicit models, with parameters constrained 
by Tevatron minimum-bias distributions. Finally, 
we apply the models
in the context of semileptonic top events at the Tevatron
and study the sensitivity of simplified top mass estimators to the
model variations.
Some previous work
leading up to this report can be found in
\cite{Sandhoff:2005jh,Sandhoff:2005}.

\section{Colour Reconnections \label{sec:colrec}}

In a first study of colour rearrangements, 
Gustafson, Pettersson, and Zerwas (GPZ) \cite{Gustafson:1988fs}
observed that, e.g.\ in hadronic $WW$ events at LEP, colour interference
effects and gluon exchanges can cause `crosstalk' between the two
$W$ systems. In the GPZ picture, the corresponding changes occurred
already at the perturbative QCD level, leading to predictions
of quite large effects.

Sj\"ostrand and Khoze (SK) \cite{Sjostrand:1993rb,Sjostrand:1993hi} 
subsequently argued against large perturbative effects 
and instead considered a scenario in which reconnections occur only
as part of the non-perturbative hadronisation phase. Starting from the 
Lund string fragmentation model \cite{Andersson:1983ia}, 
SK argued that, if two QCD strings overlap in space and time, 
there should be a finite possibility for them to fuse or cut each
other up (see e.g.\ \cite{Artru:1979ye}). 
However, since it is not known
whether the QCD vacuum more resembles a (chromomagnetic)
Type I or Type II superconductor, SK 
presented two limiting-case models, referred to
as SK-I and SK-II, respectively. Both models resulted in 
effects much smaller than for GPZ, leading to a predicted total uncertainty
on the $W$ mass from this source of $\sigma_{M_{W}}<40~\MeV$. SK
also performed a study of QCD interconnection effects in $t\tbar$ production 
\cite{Khoze:1994fu}, but only in the context of $e^+e^-$ collisions.

Subsequently, a number of alternative models have also been proposed,
most notably the ones proposed by the Lund group, based on QCD dipoles
\cite{Gustafson:1994cd,Lonnblad:1995yk,Friberg:1996xc}, and one based
on clusters by Webber \cite{Webber:1997iw}. Apart from $WW$ physics,
colour reconnections have also been proposed to model rapidity gaps
\cite{Buchmuller:1995qa,Edin:1995gi,Rathsman:1998tp,Enberg:2001vq} 
and quarkonium production \cite{Edin:1997zb}.

Experimental investigations at LEP did not find conclusive evidence  
of the effect \cite{Abbiendi:1998jb,Abbiendi:2005es,Schael:2006ns,
Abdallah:2006ve}, 
but were limited
to excluding only the most dramatic scenarios, such as GPZ and 
versions of SK-I with the recoupling strength parameter close to
unity. Furthermore, in hadron collisions
the initial state contains soft colour fields with wavelengths of
order the confinement scale. The presence of such fields,
unconstrained by LEP measurements, could impact in a
non-trivial way the formation of colour 
strings at the time of hadronisation 
\cite{Buchmuller:1995qa,Edin:1995gi}. And finally, the underlying event
produces an additional amount of displaced colour charges, translating
to a larger density of hadronising strings between the beam
remnants. It is not known to what extent the collective hadronisation
of such a system differs from a sum of independent string pieces.

As the starting point for a concrete model we take the 
description of hadron collisions developed in
\cite{Sjostrand:1987su,Sjostrand:2004pf,Sjostrand:2004ef}, as
implemented in the \textsc{Pythia} event generator
\cite{Sjostrand:2006za}. In particular, this implies that 
the underlying event is obtained by a resummation of perturbative QCD $2\to 2$ 
scatterings. The required multi-parton luminosities are obtained from
the standard 
1-parton ones, augmented with impact-parameter dependence and imposing
flavour and momentum conservation \cite{Sjostrand:2004pf}.

The system of coloured partons emerging from the
perturbative phase is dual to a set of colour-singlet QCD dipoles
\cite{Gustafson:1986db,Gustafson:1987rq}, with  
the transformation between the two being uniquely determined in the
large-$N_c$ limit (modulo a small ambiguity between the hard scattering
and underlying event initiators \cite{Sjostrand:2004pf}). 
In the absence of colour reconnections or other collective
phenomena, each such dipole translates directly to a hadronising
string piece. 

Colour reconnections can then be introduced by defining a finite probability
$\mathcal P_{\mrm{reconnect}}$ for each dipole to undergo some form of
modification before it hadronises. We shall here define the
probability for the dipole to `survive' as 
\begin{equation}
\mathcal{P}_{\mrm{keep}} = 1-\mathcal{P}_{\mrm{reconnect}} =
\prod_{i=1}^{n_{\mrm{int}}}(1-\xi_R) = 
(1-\xi_R)^{n_{\mrm{int}}}~~~,
\end{equation}
where $\xi_R$ represents an averaged probability for the dipole/string to
interact and the scaling with the number of multiple interactions
$n_{\mrm{int}}$, is intended to give a rough representation of a
scaling with the number of strings between the remnants, each of which
the dipole could have interacted with. In principle, we could have  
gone further, noting that a dipole cannot interact with
itself ($n_{\mrm{int}}\to n_{\mrm{int}}-1$), that gluon exchange
stretches two strings, rather than one, between the remnants
($n_{\mrm{int}}\to 2n_{\mrm{g}}+n_{\mrm{q}}$), the
possibility to interact with the background vacuum ($n_{\mrm{int}}\to
n_{\mrm{int}} + c$), etc.  Given the uncertainties, 
the present scaling should be a reasonable first approximation, leaving the
possibility open that future studies may require a more sophisticated
behaviour. 

By consequence, for $e^+e^-$
collisions the effective reconnection probability is simply
$\xi_R$, while it grows
from a minimum value close to $\xi_R$ for low-multiplicity
(peripheral) hadron collisions to larger values in harder, more
central collisions, such as the high-multiplicity tail of min-bias 
or top production. The model thus yields some of the
expected qualitative behaviour while leaving only a single 
free parameter, $\xi_R$, to be determined from data. 

The dipoles which do not survive define an overall colour neutral 
system of (anti-)triplet charges (a gluon 
is represented as the sum of a triplet and an antitriplet in the
$N_c\to\infty$ limit) for which a new string topology is to be
determined. 
The basis of our model is an annealing-like algorithm 
\cite{Sandhoff:2005jh} 
which attempts to minimise the total potential energy, 
as represented by  the string length measure $\Lambda$
\cite{Andersson:1983jt,Sjostrand:2002ip}, here given for 
massless partons for simplicity: 
\begin{equation}\displaystyle
\Lambda = \prod_{i=1}^N \frac{m^2_{\mrm{i}}}{M_0^2}~~~,
\end{equation}
where $i$ runs over the number of string pieces, $N$, with invariant masses
$m_{\mrm{i}}$, and $M_0$ is a constant of order the hadronisation
scale. The actual measure used by the algorithm is the
four-product of the momentum vectors of the dipole endpoints. 
We note that the minimisation of a similar measure also lies at the
heart of an earlier model for string re-interactions 
proposed by Rathsman \cite{Rathsman:1998tp}, the Generalized Area Law
(GAL), to which we plan to return in a future study.  

More aggressive models could still be
constructed, e.g.\ by reducing the risk of the annealing procedure
getting trapped in shallow local minima, but we do not consider this a
critical issue. One could also be more selective about which dipoles
to include in the annealing; here, we simply select a random set of
`active' dipoles, whereas in a more aggressive model one could have
introduced a preference for dipoles which have the most to gain in
$\Lambda$. Conversely, less
aggressive models could be motivated
by arguing that fairly long-lived resonances should be able to 
`escape' the mayhem and hadronise independently. 
Presumably, this would be particularly
relevant for colour singlet resonances, such as the $W$, and 
in a more sophisticated treatment a gradual suppression with distance
from the central hadronising region, or, more precisely, the
distance between a pair of interacting dipoles, would be expected. 
However, since 
both the top and $W$ have decay lengths of order 0.1$\ \mathrm{fm}$,
for the present we treat their decay products as fully participating in the
swapping of colours. 

Below, we investigate three variants of the algorithm; Type~0
in which the collapse of the colour field is driven by free triplets
only (gluons are sequentially attached to string systems starting from
quarks), which naturally suppresses the formation of small 
closed $gg$ string loops and is simultaneously numerically the
fastest, Type~1 in
which  $gg$ loops are suppressed by brute force, and
Type~2 in which $gg$ loops are not suppressed. The physical question 
behind this issue is, very briefly stated, whether, at the non-perturbative
transition level, gluons should be interpreted merely as representations of
transverse excitations (or `kinks') on strings whose main topology is 
defined by their quark endpoints, or whether the gluons should be allowed to
play a more independent dynamical role. Neither is likely to be the
full answer, and so by these variations we seek to explore some
measure of the associated uncertainty.

The strength parameter $\xi_R$ remains to be determined. In principle,
the constraints from LEP and diffractive processes 
would be prime candidates for this task. However,
since we are here explicitly concerned with possible breakdowns of
jet universality, the applicability of such constraints to hard inelastic
hadron collisions would be, at least, questionable. Thus, although these
connections are certainly worth exploring in more depth, we note that 
a smaller extrapolation relative to the process we are interested in 
can be obtained by simply using non-top Tevatron data. 
Here we consider minimum-bias (inelastic,
non-diffractive) events at Run II of the Tevatron, the
large-multiplicity tail of which should be fairly directly related 
to top production. In a more elaborate study, this should be extended to 
include also Drell-Yan and dijet data, particularly with $Q^2\sim
m_{\mrm{top}}^2$.  

The complex nature of hadron collisions, and in particular the
uncertainties associated with the underlying event, imply that we cannot
just correlate a single distribution with $\xi_R$ and be done with it. 
Instead, the colour reconnection strength must be determined
as part of a more general fit or `tune' to several minimum-bias 
distributions simultaneously. As a first step, we here use the charged hadron 
multiplicities, $P(N_{\mrm{ch}})$, and the mean $\pT$ as a
function of multiplicity, \pTofNch. 

The naive expectation from an uncorrelated system of strings decaying
to hadrons would be that $\avgpT$ should be independent of 
$\Nch$, and to first approximation equal to the LEP fragmentation
$\pT$ width. (With \tsc{Pythia}, the best fit for the 
non-perturbative component of this 
is $\avgpT_{\mathrm{NP}}\sim 0.36\GeV$, to give the order of magnitude.)
Already at $\mathrm{Sp\bar{p}S}$, however, and more recently
at RHIC and the Tevatron, such a constant behaviour
has been convincingly ruled out.
Currently, models which successfully
describe the  \pTofNch\ distribution, such as 
R.~Field's `Tune A' and others
\cite{Field:2005sa,Field:2005yw,Group:2006rt}, do so by  
incorporating very strong ad hoc correlations between 
final-state partons from different interactions. 
We emphasise that these correlations are not chosen at random but are
constructed to minimise the resulting string length, 
i.e.\ similarly to our models here. Thus,
although colour reconnections are not explicitly part of these
models, an implicit effect with similar consequences is still needed,
at a seemingly large 
magnitude. This observation alone serves as a significant part of the 
motivation for our study.

Given the good agreement between Tune A and Tevatron data, and given
the difficulty in obtaining the data itself, we 
constrain the new models simply by comparing to Tune
A. Tab.~\ref{tab:tunes} gives a list of 8 different \tsc{Pythia}
(v.6408) parameter
sets, which almost unavoidably combine variations of both perturbative
and non-perturbative aspects. However, we have chosen the models such
that, by comparing all of them, it should be possible to
separate the perturbative from the non-perturbative components, 
at least to a first
approximation. 

\begin{figure}
\begin{center}
\vskip-2mm \includegraphics[scale=0.39]{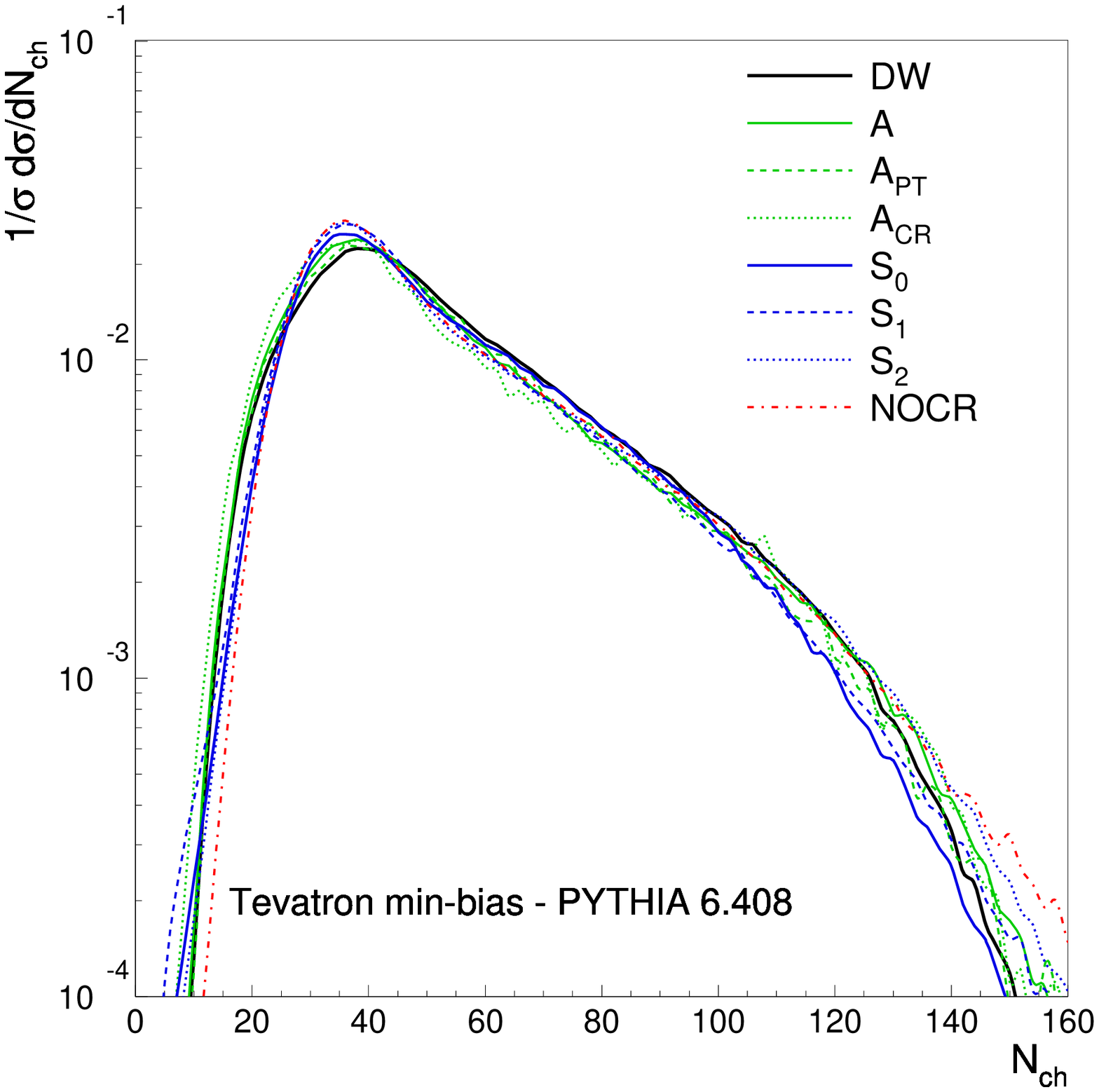}\vskip-4mm 
\includegraphics*[scale=0.39]{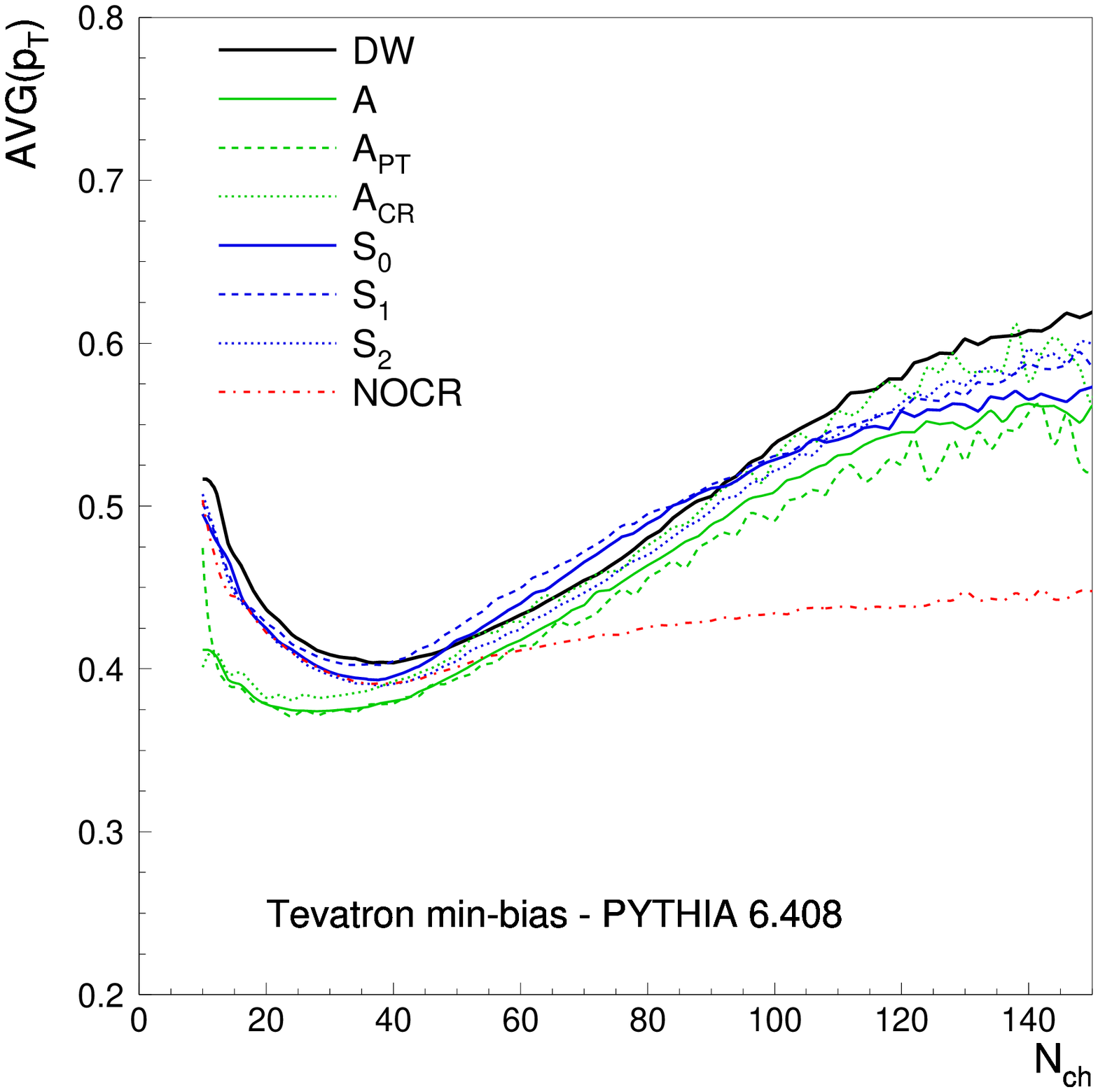}\vskip-2mm 
\caption{Comparison of the models/tunes discussed in the text. Inelastic
  non-diffractive (min-bias) events in $p\bar{p}$ collisions at
  $\sqrt{s} = 1960\GeV$. Top: Charged multiplicity distribution.
  Bottom: mean $\pT$ in GeV, as a function 
  of charged  multiplicity. The main point is not the precise
  predictions for each tune, but rather that they all roughly
  agree, with the notable exception of the \ttt{NOCR} one.\label{fig:mb}}
\end{center}\end{figure}

Tunes A and DW both pertain to the `old' UE model 
\cite{Sjostrand:1987su} and are the result of careful comparisons to CDF
data \cite{Field:2005sa,Field:2005yw,Group:2006rt}, for Tune A including
underlying event only, and for Tune DW also Drell-Yan data. We again
emphasise the large role played by non-trivial colour correlations in
these tunes, which was originally introduced to improve the fit to the
high-$\pT$ tail of hadron spectra \cite{Rick:private}. 
The A$_{\mrm{PT}}$ model is a re-tune of Tune A, with the original
virtuality-ordered final-state showers replaced by the new \pT-ordered
ones \cite{Sjostrand:2004ef} (with $\Lambda_{\mrm{QCD}}$ obtained from
a fit to \tsc{ALEPH} data \cite{rudolph}). We note that, by default, 
both choices of ordering variable incorporate matrix element merging
for hard jet radiation \cite{Norrbin:2000uu} for both top and $W$
decays. We thus expect differences in the out-of-cone effects
from hard perturbative radiation to be small. 
Similarly, A$_{\mrm{CR}}$ is again identical to Tune 
A, except that it starts from uncorrelated UE string systems and 
then applies the Type 0 colour annealing model presented
here as an afterburner. The $S_\alpha$ models pertain to the `new' interleaved
UE model \cite{Sjostrand:2004ef}, with $\pT$-ordered showers and 
Type $\alpha$ colour annealing, respectively 
--- constrained using the distributions in
Fig.~\ref{fig:mb} and, for the FSR $\Lambda_{\mrm{QCD}}$, ALEPH event shape
data \cite{rudolph}. The NOCR model is shown for reference  
only. It is the only one that does not incorporate explicit non-trivial colour
correlations, and even then care has been taken to exploit the
initial-state colour ambiguity mentioned above to the fullest.

\begin{table*}[t]
{\footnotesize\begin{tabular}{ll|cccc|cccc|}
Parameter (\tsc{Pythia} v.6408+)& 
 &  DW & A & A$_\mrm{PT}$ & A$_\mrm{CR}$ & S$_0$ & S$_1$ &
 S$_2$ & \footnotesize{NOCR} \\\hline\hline
% FSR $Q^2_\mrm{max}$ factor (non-s-channel) &  PARP(71)&4.0& 4.0&4.0&4.0&\\
UE model & MSTP(81) &\multicolumn{4}{c|}{
 1 (`old' \cite{Sjostrand:1987su})}&\multicolumn{4}{c|}{21 (`new'
 \cite{Sjostrand:2004ef})}\\ 
UE infrared regularisation scale (at $\sqrt{s}=1800\GeV$)& 
 PARP(82)&1.9&2.0 &2.1&2.0&1.85&2.1&1.9&2.05\\
\ \ -''-, scaling power with $\sqrt{s}$ & PARP(90) &
\multicolumn{4}{c|}{0.25 (`fast')}&\multicolumn{4}{c|}{0.16 (`slow')} \\ 
UE hadron transverse mass distribution &  MSTP(82)&\multicolumn{4}{c|}{4
 (`double Gaussian')}&\multicolumn{4}{c|}{5 (`ExpOfPow')}\\
 \ \ -''- parameter 1 & PARP(83) &\multicolumn{4}{c|}{0.5}
& 1.6&1.4&1.2&1.8 \\ 
 \ \ -''- parameter 2 & PARP(84) &\multicolumn{4}{c|}{0.4} 
& \multicolumn{4}{c|}{n/a}\\
 UE total $gg$ fraction &
 PARP(86) &1.0&
 0.95&0.95&0.66&\multicolumn{4}{c|}{n/a}
 \\ \hline
 ISR infrared cutoff& PARP(62)&1.25&1.0&1.0&1.0&
 \multicolumn{4}{c|}{($\equiv$ PARP(82) )}\\   
 ISR renormalisation scale prefactor & PARP(64)&0.2& 1.0&1.0&1.0&\multicolumn{4}{c|}{1.0}\\  
 ISR $Q^2_{\mrm{max}}$ factor & PARP(67)&2.5&
 4.0&4.0&4.0&\multicolumn{4}{c|}{n/a}\\
 ISR infrared regularisation scheme & MSTP(70) & 
 \multicolumn{4}{c|}{n/a} & 2 & 0 & 2 & 2\\ 
 ISR FSR off ISR scheme & MSTP(72) & 
 \multicolumn{4}{c|}{n/a} & 0 & 1 & 0 & 0\\ 
\hline
 FSR model &  MSTJ(41)& 2 &
 2 & 12 & 2 &\multicolumn{4}{c|}{(\pT-ordered)}\\ 
 FSR $\Lambda_{\mrm{QCD}}$ & PARJ(81) & 0.29 & 0.29 & 0.14 & 0.29 & 
 \multicolumn{4}{c|}{0.14}\\ 
\hline
 BR colour scheme & MSTP(89)
 &\multicolumn{4}{c|}{n/a}&1&1&1&2 \\
 BR composite $x$ enhancement factor & PARP(79)
 &\multicolumn{4}{c|}{n/a}&2&2&2&3 \\
 BR primordial $k_T$ width $\langle\vert k_T\vert\rangle$
 &   PARP(91)&2.1&  1.0
 &1.0&1.0&\multicolumn{4}{c|}{n/a} \\ 
 BR primordial $k_T$ UV cutoff
 &   PARP(93)&15.0&  5.0
 &5.0&5.0&\multicolumn{4}{c|}{5.0} \\ \hline
 CR model & MSTP(95) &\multicolumn{3}{c}{n/a}&
 6 & 6 & 2 & 4 & 1 \\
 CR strength $\xi_R$ & PARP(78) &\multicolumn{3}{c}{n/a}&
 0.25 & 0.2 & 0.35 & 0.15& 0.0\\
 CR $gg$ fraction (old model) & PARP(85) & 1.0 & 0.9 & 0.9 &
 0.0&\multicolumn{4}{c|}{n/a}  \\\hline\hline
 \end{tabular}}
\caption{\textsc{Pythia} parameters  \cite{Sjostrand:2006za}, 
divided into a few main categories: UE (underlying event), ISR
(initial state radiation),  
  FSR (final-state radiation), BR
  (beam remnants), and CR (colour
  reconnections). The UE reference
  energy for all models is \parm{PARP(89)=1800}\GeV, and all
  dimensionful parameters are given in units of \GeV. \ttt{MSTP(95)=2,4,6}
  corresponds to CR types 1, 2, and 0, respectively, in the text.
\label{tab:tunes}}
\end{table*}
In Fig.~\ref{fig:mb}, the 8 models in Tab.~\ref{tab:tunes} are
compared on the $N_{\mrm{ch}}$ and \pTofNch\ distributions. 
A good description of the charged multiplicity distribution is
obtained in all cases. Simultaneous good agreement with  \pTofNch\ is
only obtained for the models incorporating non-trivial colour
correlations -- the notable exception being the NOCR 
model which exhibits the close-to-constant behaviour of uncorrelated
string decays discussed above. We interpret this behaviour as representing a
concrete example of a data-driven motivation to develop ideas of
hadronisation beyond the current cluster/string models, which have
remained essentially frozen since the LEP era. This is not to imply
that these models are intrinsically unsatisfactory or that 
hadronisation at LEP should be perceived as being a completely
separate story, but only that 1) we expect that the many recent
improvements on the perturbative side imply that there is  
less `wriggle' room for the non-perturbative physics, and hence the
latter could presumably be better constrained today than a decade ago,
and 2) the 
size of possible differences between non-perturbative effects in different
environments should be more fully explored, where current models are
normally limited to the assumption of jet universality. 

Returning to the question at hand, 
the remaining models in Tab.~\ref{tab:tunes} all describe the two
distributions in Fig.~\ref{fig:mb} within an acceptable margin, 
at least as gauged by the spread between the 
two more elaborate Tevatron tunes, A and DW. 

\section{The Top Mass at the Tevatron}

Assuming the in situ extrapolation discussed
above to be at least moderately reliable, we now apply the same 
models and parameters in the context of top production at the
Tevatron. More specifically, we concentrate on semileptonic $t\bar{t}$
events, i.e.\ $t\bar{t}\to b\bar{b}q\bar{q}\ell\nu$. 
As a first step towards estimating the sensitivity of
experimental top mass observables, we consider the impact on 
a simplified measurement which roughly approximates the  
key ingredients used in current Tevatron analyses. 
As already mentioned, it is not possible to cleanly separate the CR
effect from other sources of variation in these models. 
Nonetheless, by grouping models with 
similar parton showers and studying both the variations within and
between the groups, we are able to make some headway.
Differences between the results obtained using the same parton shower
but different CR models for the same generated top mass
may then be interpreted as a first estimate of the uncertainty on
$m_\mathrm{top}$ due to 
genuine non-perturbative effects, while the uncertainty between 
groups with different shower models is interpreted as having a
perturbative origin. 

\subsection{Real Top Mass Measurement}
To set the stage for the subsequent analysis, let us first summarise
the methods used in actual top mass reconstructions at the Tevatron,
focusing on the ingredients they employ to go from detector level events to a
reconstructed top mass.  
The relevant measurements performed in the semileptonic channel by the CDF and D\O\ 
collaborations~\cite{Abulencia:2005aj,Abulencia:2005pe,Abazov:2006bd,d0template,Abazov:2007rk}
use one of three methods.  

The \em template method \em compares the distribution of kinematically 
reconstructed top mass values to templates obtained for various 
nominal top mass
values from simulation with full detector description (including background). 
The \em matrix element method \em computes the event-by-event likelihood that the observed
kinematic configurations stem from %top pair events of with a given
events of a given top mass. Maximising the total likelihood of the
observed sample yields the final result.
Finally, the \em ideogram method \em reconstructs top mass values in each
event and then builds the likelihood of observing that value with the given
resolution as a function of the true value.  
Again, maximising the total likelihood of the observed sample
yields the final result.

All three methods are based on so-called reconstructed physics objects,
i.e.\ jets, identified charged leptons, and missing transverse energy, 
which fulfil certain selection criteria. 
For the methods with explicit reconstruction of
the top mass an  assignment of jets to one of the two top quarks has to be
made. Often constrained fits are used to improve the experimental resolution by
requiring that the two reconstructed top masses are equal and that the known
$W$-mass is reconstructed in the decay of each of the top quarks.
In this case an assignment of each jet to the $W$-boson or $b$-quark
within each top decay product is also required. 

The methods are calibrated, i.e.\ the performance of a given
implementation is measured on fully simulated events. Any deviation 
of the reconstructed from the generated top masses is corrected for; 
this implies that the top mass
measurement is limited by the precision of the simulation used for
the calibration. 

Since summer 2005 all methods have been extended to tackle the dominant
experimental systematic uncertainty, the jet energy scale (JES), by simultaneously
fitting the top mass and the jet energy scale, with the
additional constraint that the reconstructed mass of the 
hadronically decaying $W$ present in each event should be consistent with
the known $m_W$. 

The main features of a top mass measurement are thus utilisation of
reconstructed physics objects, assignment of each jet to a specific top
decay product, correction for an overall JES factor, 
and calibration of the method. We shall now construct a simplified top
mass estimator which embodies these four ingredients. 

\subsection{Toy Top Mass Measurement for Generator Level}

To obtain an estimate of the influence of the above CR/UE models on top
mass measurements a toy mass measurement on events generated by
\tsc{Pythia} was implemented. The study is performed at the generator
level, here meaning after hadronisation and hadron decays but without
detector simulation. 

In this simplified analysis electrons and neutrinos are 
`identified' by looking at the generator truth. 
Jets are reconstructed on final-state particles using cone jets of 
$\Delta R=0.5$~\cite{ktjet,plano} 
%or $k_T$-jets with $d_\mathrm{cut}=150\GeV^2$~\cite{ktjet} 
and are required to have a $p_T>15\GeV$. Only semileptonic events with exactly 
four reconstructed jets are considered for further analysis.

Jets are assigned to the top decay parton with the lowest $\Delta R$
distance. Events without a unique one-to-one assignment are discarded.
In the samples used here, between $25\%$ and $27\%$ of the semileptonic
events fulfil these requirements. 

The top mass is now computed event by event 
from the sum of the four-momenta of the three
jets from the hadronically decaying top in the event.
The top mass for the full sample is then obtained by fitting a Gaussian to
the distribution of reconstructed masses. To minimise the importance of
the tails of the distribution, the fit range is restricted to a window
of $\pm 15\GeV$ around the top mass. This width corresponds 
approximately to the  experimental resolution observed in current measurements.
The fit is iterated until it settles on a range symmetric around the final
result, $m_\mathrm{top}^\mathrm{fit}$. 

The same method can also be applied for measuring the $W$-mass from the two
jets assigned to the partons from the $W$ decay. 
The resulting $W$-mass measurement can be used to compute a JES correction
factor: $s_{\mathrm{JES}}=80.4\GeV / m_W$. The 
fitted top mass above can then be corrected by this factor, to produce a
JES-scaled top mass value,\linebreak
$m_\mathrm{top}^\mathrm{scaled}=s_{\mathrm{JES}}\cdot
m_\mathrm{top}^\mathrm{fit}$. 

This full procedure is repeated for several different values of the
generated top mass, between 165 and 185\GeV, in steps of 1\GeV. 
Both the scaled and unscaled fitted top masses exhibit a completely
linear dependence on the input top mass. Observed slopes are
consistent with 1 at the 
2\% level. This indicates that the fit
procedure is indeed stable and has the desired dependence on the
physical quantity, cf. Fig.~\ref{fig:calib}.

\subsection{Differences from different Models}

\begin{figure}
\vspace*{-3mm}
\begin{center}
{
\unitlength=1mm
\begin{picture}(80,80)
\put(17,40){\includegraphics[width=3cm]{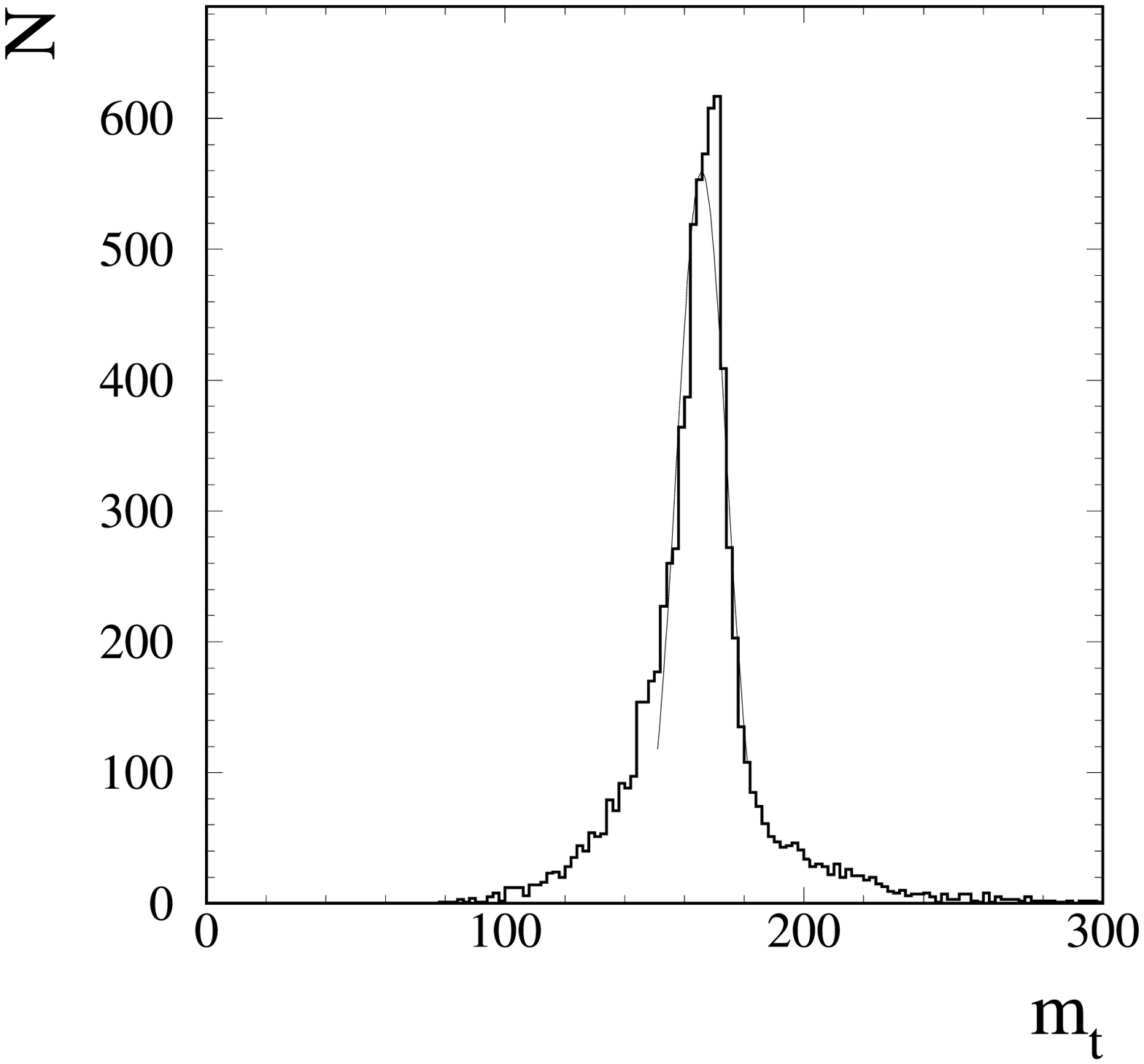}}
\put(0,0){\includegraphics[width=0.45\textwidth,clip,trim=10mm 50mm 0mm 4mm]{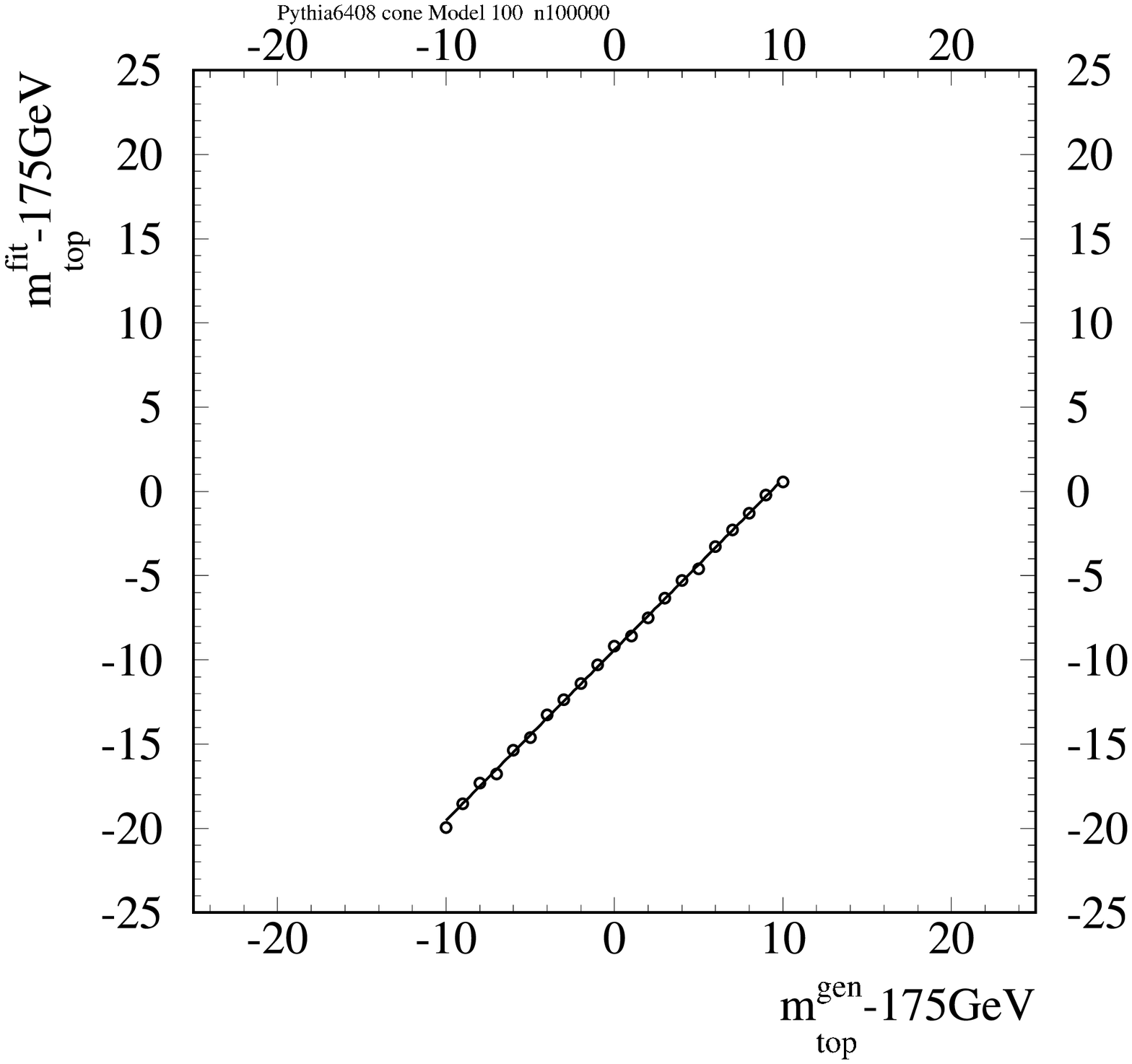}}
\end{picture}
}\vspace*{-5mm}
\end{center}
\caption{\label{fig:calib}%
 Calibration curve obtained for Tune A, before JES rescaling. 
A similar plot was
  made for each model in Tab.~\ref{tab:tunes} and their relative
  offsets compared, both before and after JES rescaling. 
The inset shows the Gaussian
  fit to the distribution reconstructed top masses from the hadronic
  event side for the specific point
  $m_\mathrm{top}^{\mathrm{gen}}=175\GeV$. \label{fig:offset1} 
}
\end{figure}

In real mass measurements, the offset and slope of the straight line
that describes the reconstructed vs.\ the generated top mass 
is used to calibrate the given top mass procedure. 
However, this calibration must necessarily use only one specific CR/UE model.
By virtue of the tuning we performed, there is a genuine ambiguity of
which model to choose. 
Differences between the individual model calibrations therefore lead 
to uncertainties on the top mass
results.  As an example, Fig.~\ref{fig:offset1} shows the calibration curve
obtained for Tune A before JES rescaling. 

%\framebox
\begin{figure}
\begin{center}
\vspace*{-2mm}
\unitlength=1mm
\begin{picture}(0,0)
\put(9.5,62){$\displaystyle\overbrace{\hspace*{10mm}}^{\Delta m_\mathrm{top}^\mathrm{fit}}$}
\put(60,62){$\displaystyle\overbrace{\hspace*{14mm}}^{\Delta m_\mathrm{top}^\mathrm{scaled}}$}
\end{picture}
\includegraphics[width=0.45\textwidth]{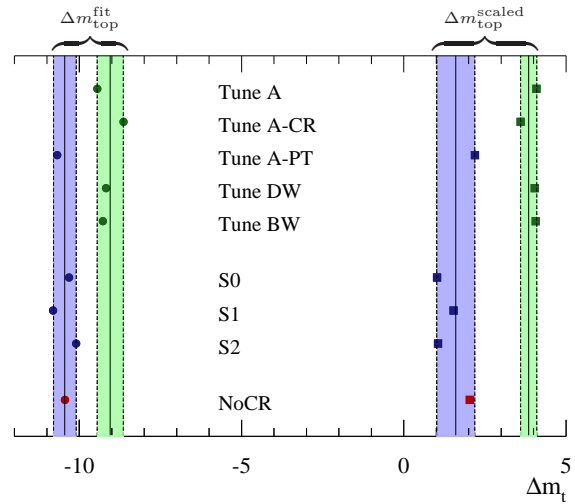}
\vspace*{-7mm}
\end{center}
\caption{\label{fig:offset2}%
Comparison of calibration offsets obtained for each model, in GeV, here
including an additional parameter set, `BW', from Rick Field. 
On the left  are the results obtained before $\mathrm{JES}$ 
rescaling (dots) and on the right after rescaling (squares). The
coloured bands group models
with the same final-state shower (green: virtuality-ordered, blue:
\pT-ordered). The statistical
precision due to the finite number of generated events is at the 
$0.1\GeV$ level. 
}
\end{figure}

Fig.~\ref{fig:offset2} summarises our central results. It shows the offsets
before (left) and after (right) scaling with the $\mathrm{JES}$
correction factor. 
The offsets, $\Delta m_\mathrm{top}$, for each model are obtained from a straight line fit to the
calibration curve evaluated at $m_\mathrm{top}^\mathrm{gen}=175\GeV$, 
and have statistical precisions (determined from
the spread of the data points) of $\sim0.1\GeV$.
%The input top mass for this plot was 175\GeV, but
%due to the high linearity of the calibration curves the relative
%offsets are approximately constant over the range of $m_t$ we have
%considered. 

The top masses in the uncorrected fits (dots, left column)
come out somewhat lower than the input mass, principally due to
out-of-cone corrections. Including the JES correction, i.e.\ scaling all
jets by the factor necessary to get the right hadronic $W$ mass, the
points move to the right (squares), even to the point of
over-correcting the top mass. 

Again, our central point is that, while for any particular model 
a further, constant offset would be sufficient to calibrate the measurement 
to coincide with the input mass, the spread between models cannot be
dealt with in this way. It is the 
ambiguity coming from not knowing \em which \em offset value to
correct for that we interpret as the uncertainty on the top mass. 

It therefore seems significant that the various models 
exhibit differences of about $\pm 1.1\GeV$ and $\pm 1.5\GeV$ for the offsets of
$m_\mathrm{top}^\mathrm{fit}$ and  $m_\mathrm{top}^\mathrm{scaled}$, respectively. 
Explicit checks varying both the fit range and fit function produced
variations no larger than $\sim 20\%$ in these numbers, hence at this
level the effect appears genuine. Without additional
constraints from data, it translates directly into an   
uncertainty on the reconstructed top mass. 

To extricate the genuinely non-perturbative part of this, 
we note that the models
fall into two broad classes: those that utilise the ``old''
virtuality-ordered final-state 
parton shower and those that utilise the ``new'' \pT-ordered one. The largest
component of the difference is \emph{between} these two classes,
hinting at a perturbative origin for most of it, which, at least 
to some extent, should already be present in Tevatron analyses via the
\tsc{Pythia}-\tsc{Herwig} systematic. 

Within each class, we still observe differences roughly of order
$\pm0.5\GeV$ on the top mass, which we are more confident in
assigning a non-perturbative origin. Note, however, that this still
lumps genuine CR effects together with other infrared ambiguities, 
such as infrared regularisation and renormalisation procedures, 
the treatment of beam remnants, etc.%

Real mass analyses may have a different sensitivity to the
model differences. 
The size of the effect in this first study, however, 
suggest a need for further in depth analyses.
If the sensitivity we observe here is confirmed for real mass analyses, 
we hope the question may be turned
around, and that in situ measurements can be used to gain further
information about the interesting physics effects that may be present. 

\section{Summary and Conclusion}

We have presented a set of new, universally applicable models to study 
colour reconnection (CR) effects in hadronic final states. 
The models are 
based on hadronising strings and apply an annealing-like algorithm to 
minimise a measure of the classical potential energy, with a freely
varying strength parameter running from zero to unity. A scaling is
included such that the survival
probability of a given string piece decreases
as a function of the number of perturbative scatterings in the
underlying event. The models are implemented in a publicly available
version (v.6408+) of the \tsc{Pythia} event generator. 

To constrain the CR strength parameter we have used 
Tevatron minimum-bias distributions, specifically 
$P(N_{\mrm{ch}})$ and \pTofNch. Taking the
results obtained with the CDF `Tune A' as a benchmark, we present several
alternative parameter sets exploring the possible combinations of showers,
underlying-event 
modelling, and colour reconnections. As a further, data-driven 
motivation, we argue that current underlying-event
descriptions, including `Tune A', already include strong non-trivial
colour correlations. 

As a first application, we have investigated the influence of changing
the underlying physics model, including CR, UE, and shower effects, on
a range of simplified Tevatron top mass measurements. The models we
consider exhibit individual variations of about $\pm1.5\GeV$ on the
reconstructed top mass. 
{
While this is comparable to systematic uncertainties quoted for
present top
mass measurements, it has so far only partly been considered in 
the current analyses.

Of the total variation we attribute about $\pm 1\GeV$ to
perturbative effects and about $\pm 0.5\GeV$ to non-perturbative
sources.
}

Our conclusion for the present is thus twofold: firstly, 
colour reconnections in hadron collisions appear to be a both
experimentally and theoretically motivated possibility, one which
should be explored as part of developing a more detailed picture of
hadron collisions. 
Secondly, it appears that non-perturbative uncertainties, among which
colour reconnections hold a prominent place, are likely to be relevant in the 
drive towards sub-\GeV\ uncertainties on the top mass at the Tevatron. 

It is important to now verify the size of the observed uncertainties
in real mass measurements and
to increase the amount of non-top data used to constrain the models; Drell-Yan
production, and in particular its high-mass tail, is likely to be 
useful in reducing the initial-state shower ambiguities, while
the infrared effects could be further probed by expanding on the number
of minimum-bias distributions, as well as including underlying-event 
studies in dijet and Drell-Yan production, again in particular when
$Q\sim m_{\mrm{top}}$ where the required extrapolation is presumably minimal. 
The connection with the LEP data (and, possibly, diffractive physics) 
should also be explored, although as we have noted the
assumption of jet universality should probably not be treated as inviolate in
this context. 

On the theoretical side, we hope that the arguments we have presented
will stimulate curiosity, and eventually activity, in this now
somewhat dormant field. Along the intersection of the two communities, it
would be interesting to explore alternative measurement
strategies and, as a last resort, a combined tuning and top mass fit.
A final follow-up we envision is to
extrapolate in energy to evaluate the impact on precision studies at
the LHC. 

\subsubsection*{Acknowledgements:}
We are grateful to F.~Canelli, F.~Fiedler, K.~Hamacher, J.~Rathsman,
M.~Sandhoff, and T.~Sj\"ostrand 
for enlightening discussions and valuable comments on the draft. P.S.\ is
supported by  
Fermi Research Alliance, LLC, under Contract No.\
DE-AC02-07CH11359 with the United States Department of Energy

\bibliography{colrec-epjc-rev}

\end{document}